\documentclass[reprint,amsmath,amssymb, aps, prl,floatfix,superscriptaddress]{revtex4-1}
\usepackage{graphicx}
\usepackage{bm}

\begin{document}

\setlength{\unitlength}{1cm}
\preprint{APS/123-QED}

\title{Topological Stabilization and Dynamics of Self-propelling Nematic Shells}%
 \author{Babak Vajdi Hokmabad}%
\thanks{BVH and KAB contributed equally}%
\affiliation{Max Planck Institute for Dynamics and Self-Organization, Am Fa\ss{}berg 17, 37077 G\"ottingen}
\affiliation{Institute for the Dynamics of Complex Systems, Georg August Universit\"at G\"ottingen}
 \author{Kyle A. Baldwin}%
\thanks{BVH and KAB contributed equally}%
\affiliation{Max Planck Institute for Dynamics and Self-Organization, Am Fa\ss{}berg 17, 37077 G\"ottingen}
 \affiliation{School of Science and Technology, Nottingham Trent University, Nottingham, NG11 8NS, United Kingdom}
\author{Carsten Kr\"uger}%
\affiliation{Max Planck Institute for Dynamics and Self-Organization, Am Fa\ss{}berg 17, 37077 G\"ottingen}
\author{Christian Bahr}%
\affiliation{Max Planck Institute for Dynamics and Self-Organization, Am Fa\ss{}berg 17, 37077 G\"ottingen}
 \author{Corinna C. Maass}%
 \email{corinna.maass@ds.mpg.de}%
\affiliation{Max Planck Institute for Dynamics and Self-Organization, Am Fa\ss{}berg 17, 37077 G\"ottingen}
\date{\today}%

\newcommand{\beginsupplement}{%
        \setcounter{table}{0}
        \renewcommand{\thetable}{S\arabic{table}}%
        \setcounter{figure}{0}
        \renewcommand{\thefigure}{S\arabic{figure}}%
     }

\begin{abstract}
Liquid shells (\emph{e.g.} double emulsions, vesicles \emph{etc.}) are susceptible to interfacial instability and rupturing when driven out of mechanical equilibrium. This poses a significant challenge for the design of liquid shell based micro-machines, where the goal is to maintain stability and dynamical control in combination with motility. Here we present our solution to this problem with controllable self-propelling liquid shells, which we have stabilized using the soft topological constraints imposed by a nematogen oil. We demonstrate, through experiments and simulations, that anisotropic elasticity can counterbalance the destabilizing effect of viscous drag induced by shell motility, and inhibit rupturing. We analyze their propulsion dynamics, and identify a peculiar meandering behavior driven by a combination of topological and chemical spontaneously broken symmetries. Based on our understanding of these symmetry breaking mechanisms, we provide routes to control shell motion via topology, chemical signaling and hydrodynamic interactions.
\end{abstract}

\maketitle

The capability to produce controllable, actively self-propelling microcapsules would present a leap forward in the development of artificial cells, microreactors, and microsensors. 
Inactive microcapsules have been developed in the form of double emulsions (droplet shells), which have been applied as, \emph{e.g.}, reactive microcontainers~\cite{lorenceau2005_generation}, synthetic cell membranes~\cite{petit2016_vesicles-on-a-chip}, food and drug capsules~\cite{augustin2009_nano,kim2011_double-emulsion}, optical devices~\cite{chen2014_photoresponsive,nagelberg2017_reconfigurable,zarzar2015_dynamically}, and biotic sensors~\cite{zhang2017_janus}. 
However, these highly structured compound droplets are usually non-motile, and any actuation that displaces their liquid cores makes them susceptible to shell rupture if the interfaces of the nested compartments can coalesce. 
Shells stabilized by current techniques such as vesicles, capsids or polymersomes possess immobile interfaces which impede self-actuation. 
Hence, engineering such motile systems requires further complexities in design and fabrication~\cite{kumar2018_enzyme-powered,singh2017_microemulsion-based,joseph2017_chemotactic}. 

To survive motility, any liquid shell with mobile interfaces requires a stabilizing force to counter the destabilizing swimming dynamics. 
Here we present a new approach to the problem of combining encapsulation with autonomous motility, by using nematic active double emulsions, where anisotropic micellar solubilisation induces motility, and nemato-elasticity of the shell provides stability without requiring further complexities in the design. 
Through experiments and simulation of the elastic energy in the liquid crystal shell, we show that active shells are stable only in the nematic state. 
We demonstrate that the shell dynamics are dictated by anisotropic self-generated chemical fields, broken topological symmetries, and hydrodynamic interactions, and that by tuning these factors we can control and direct their motion, providing avenues for applications in transport, guidance and targeted release. Our framework provides a bottom-up approach for developing functional micro-machines using established physicochemical mechanisms.

Our active double emulsion system is comprised of water-in-oil-in-water droplet shells, where self-propulsion is induced via solubilisation in a micellar surfactant solution, and maintained by a self-sustaining surface tension gradient in the external oil-water interface~\cite{herminghaus2014_interfacial, maass2016_swimming, izri2014_self-propulsion}. 
Swimming droplets shed persistent trails of oil-filled micelles, from which they are subsequently repelled~\cite{jin2017_chemotaxis}. 
We use the nematogen 5CB as the oil phase, and solutions of the anionic surfactant TTAB as the aqueous phases, where the internal core droplet is submicellar ($c=0.75\,\mathrm{CMC}$), and the external swimming medium is supramicellar ($c>30\,\mathrm{CMC}$). 
We mass-produce highly monodisperse oil droplet shells using consecutive microfluidic cross-junctions in flow-focusing configuration ~\cite{SM,petit2016_vesicles-on-a-chip}. 

\begin{figure}
\includegraphics[width=\columnwidth]{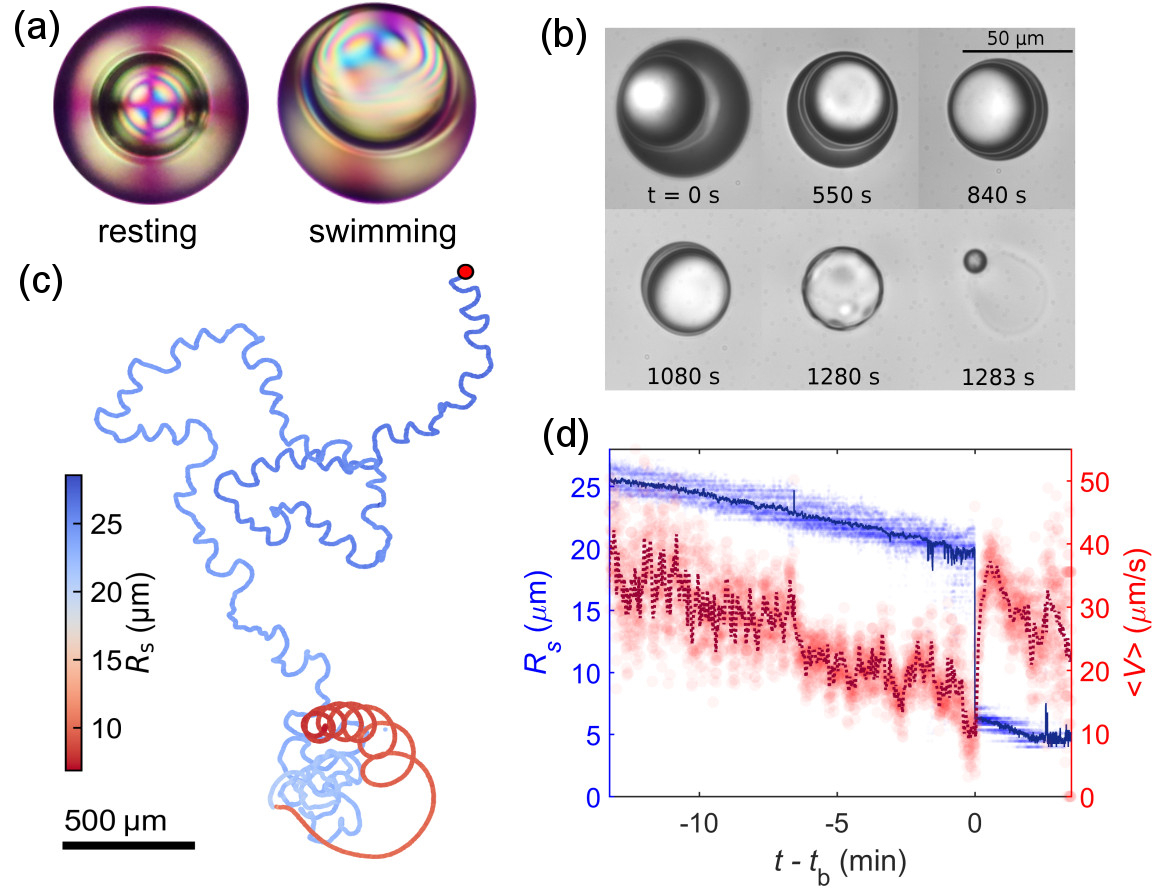}
\caption{Life stages of active shells.\
(a)\ Polarised images of resting and swimming shell.\ 
(b)\ Video stills of droplet life stages.\
(c)\ Example trajectory coloured by shell radius $R_{s}$, recorded over 40\,min. The sudden change of colour from blue to red corresponds to the burst moment. 
(d) Average speed $V$ (dotted, red) and radius $R_{s}$ (solid, blue) for 13 shells, time $t$ relative to bursting time $t_{b}$ (scatter plots: values for all experiments). \label{fig:lifestages}}
\end{figure}

\begin{figure*}
\centering\includegraphics[width=.9\textwidth]{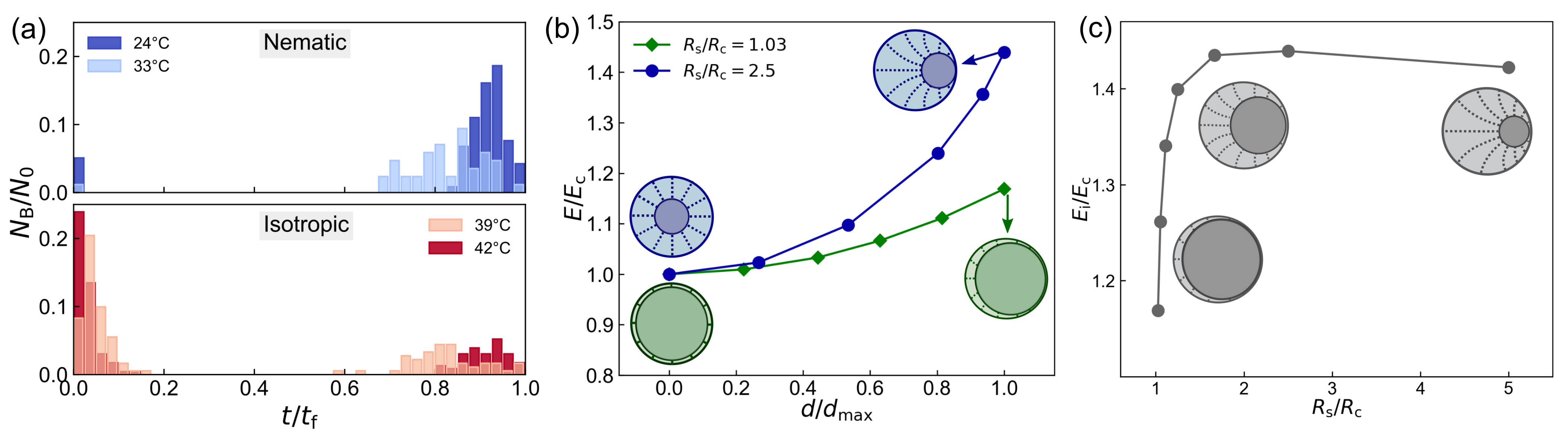}
\caption{Stabilization of the active shells.
(a)\ Burst statistics (number of bursts $N_{B}$ normalised to initial number of shells $N_0$) for shells below and above the clearing point, plotted against time $t$ normalised to final time $t_{f}$ with no remaining shells ($t_{f}(24^\circ$C)$\approx45$\,min., $t_{f}(39^\circ$C$)\approx10$\,min., initial $R_{s}=30\,\mu{m}, R_{c}=18\,\mu{m}$)\
(b)\ Elastic energy $E/E_{c}$ (${E_{c}}$: core at centre of droplet) as a function of core displacement $d/d_{\scriptscriptstyle{max}}$, with $d_{\scriptscriptstyle{max}} = R_{s} - R_{c} - 100$\,nm, for two different ratios $R_{s}/R_{c}$.\
(c)\ Ratio of elastic energies ${E_{i}}/{E_{c}}$ (${E_{i}}$: core droplet close to outer shell interface) against ratio of radii $R_{s}/R_{c}$, using $R_{c} = 25\,\mathrm{\mu{m}}$ and $R_{s}$ shrinking from 125\,$\mu$m to 25.6\,$\mu$m. Schematics are illustrative, illustrated radius values do not directly correspond to the simulation parameters.
\label{fig:stability}}
\end{figure*}

Despite the displacement of the aqueous core towards the shell boundary (Fig.~\ref{fig:lifestages}\,a), the shells self-propel stably and reproducibly for long times, dissolving down to thin shells with a minimum stable shell/core radii fraction of $R_{s}/R_{c}\approx1.05$. 
The life stages of these self-propelling shells (Fig.~\ref{fig:lifestages}\,b and c) fall into three regimes: (I) \textit{`Shark-fin' meandering}. At early times, the core is small compared to the shell diameter (Fig.~\ref{fig:lifestages}\,b, top) and is deflected considerably from the polar axis of the shell, resulting in a meandering instability. 
(II) \textit{Thin shells.} As the shell thins, eventually there is little room for significant asymmetry in the shell-core arrangement (Fig.~\ref{fig:lifestages}\,b, bottom). 
During this stage, the motion grows noisy while the speed decreases, until propulsion stops. 
(III) \textit{Single Emulsion}. On reaching a critical minimum thickness, the shell bursts, reconstituting into a single oil droplet. 
From a comparison of pre- and post-burst radii, we estimate the average shell thickness at this point is less than $1~{\mu m}$ (Fig.~\ref{fig:lifestages}\,d). 
The droplet then propels with an undisturbed internal convection, leading to a sudden increase in speed (Fig.~\ref{fig:lifestages}\,d), and a curling motion as observed in nematic single emulsions~\cite{kruger2016_curling}(Fig.~\ref{fig:lifestages}\,c). 

In contrast to these reproducible stages in nematic shells, we find that under otherwise identical conditions, shells made from isotropic oils (CB15 or 5CB/BPD, see~\cite{SM}) burst significantly earlier. 
Fig.~\ref{fig:stability}\,a shows burst statistics for 5CB shells, where below the clearing point ($T<34.5\,^\circ${C}, nematic) shells survive for long times, whereas above the clearing point ($T>34.5\,^\circ$C, isotropic), most droplets do not reach the thin shell stage. 

We attribute the shell stability to a nemato-elastic energy barrier: 5CB molecules arrange to minimise the elastic energy associated with the deviations from a uniform director field imposed by the boundary conditions (here homeotropic anchoring~\cite{brake2003_effect}). 
In a resting shell, this causes a radially symmetric arrangement of the director field~\cite{lopez-leon2011_drops} with the aqueous core at the centre. 
In a moving shell the internal flow drives the core off-centre: the director field is therefore distorted both by the displacement of the core and the flow field, such that the stored elastic energy is increased. 

To estimate the competing forces we numerically simulated the director field inside the shell and calculated the elastic energy $E$ stored in a resting shell with a core displaced by a distance $d$. We applied a common numeric minimisation technique~\cite{mori1999_multidimensional, ravnik2009_landau-de} based on the $\mathbf{Q}$ tensor representation~\cite{degennes1971_short} of the nematic director field~\cite{SM}. The tensor elements of a uniaxial nematic with scalar order parameter $S$ and local director $\textbf{n}$ are given by:
\begin{equation}
Q_{jk} = \frac{S}{2} \left( 3 \textbf{n}_j \textbf{n}_k - \delta_{jk} \right).
\end{equation}
Since topological defects are not present in the director field of our shells, we neglected a variation of the magnitude of $S$ and assumed a constant value $S=1$. For the calculation of the elastic energy density $f_e$ we used the one-constant approximation of the nematic elasticity, \emph{i.e.}, $K_\mathit{splay} = K_\mathit{twist} = K_\mathit{bend} = K$. Then, $f_e$ is obtained as:
\begin{equation}
f_e = \frac{K}{9} Q_{jk,l} Q_{jk,l}
\end{equation} 
where, $Q_{jk,l} = \partial_l Q_{jk}$. The total elastic energy $E$ is then calculated by integration over the shell volume $\Omega$:
\begin{equation}
E = \int_\Omega f_e \mathrm{d} \Omega.
\end{equation}
As shown in (Fig.~\ref{fig:stability}\,b), we find that $E$ increases by a factor of $E_{i}/E_{c}\approx 1.4$ when the core droplet is located at the outer interface ($E=E_{i}, d=d_{max}$), as compared to the centred configuration ($E=E_{c}$). 
Remarkably, we find only a minor dependence on the thickness of the nematic shell. 
$E_{i}/E_{c}$ drops significantly towards unity only for $R_{s}/R_{c} < 1.1$ (Fig.~\ref{fig:stability}\,c), \emph{i.e.}, the elastic energy barrier vanishes only in the limit of zero shell thickness.

\begin{figure*}
\includegraphics[width=.9\textwidth]{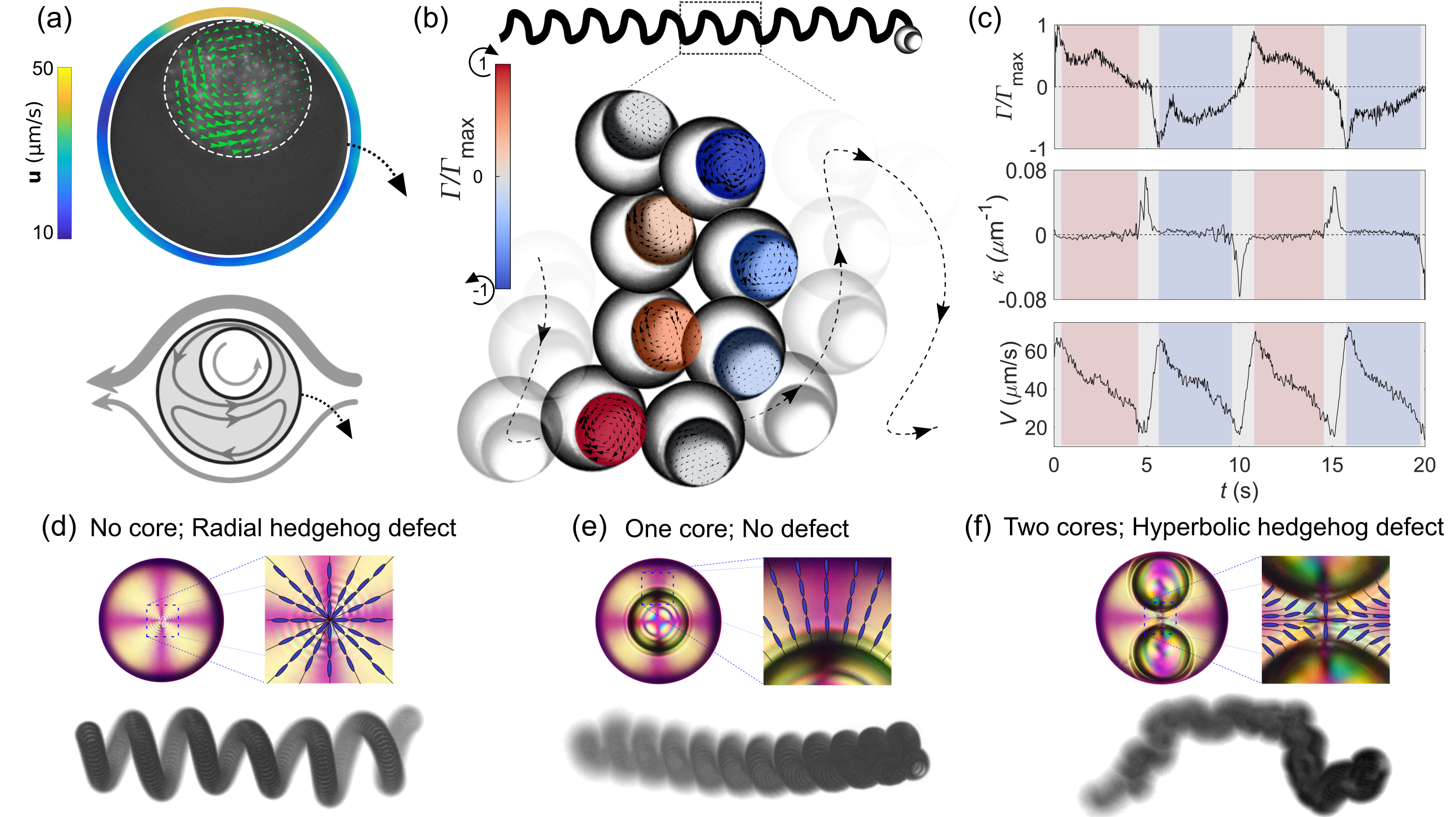}%

\caption{\label{fig:coreorient} Swimming behaviour:\ 
(a) Top: PIV data for flow at the outer interface and in the core. Bottom: schematic of core and flow arrangement.\ 
(b)\ Top: `shark-fin' meandering of swimming trajectory over 2 minutes ($R_{s}=36\,\mu$m); the shell switches periodically between clockwise and anticlockwise turns. Bottom: zoomed in view of the shell-core alignment with the swimming trajectory, with superimposed core flow fields and colour-coded circulation.\ 
(c)\ Core circulation $\Gamma/\Gamma_{\scriptscriptstyle{max}}$, local curvature $\kappa$, and shell speed $V$ versus time over four meandering periods.\
(d), (e), and (f) show the nematic structure (top, polarised images of the droplets at rest with overlaid director field schematic) and 3D swimming trajectories (bottom, multiple exposure micrograph captured over 60 s) for a no core droplet (d), a single core droplet (e) and a two core droplet (f).
See \cite{SM} movie S8.}
\end{figure*}

We calculate the elastic force $F_{e}=\partial E/\partial d$ acting on a core displaced to the boundary of a 5CB shell~\cite{karat1977_elasticity, chmielewski1986_viscosity} to be of the order of $\approx100$\,pN. 
This is equivalent to the Stokes drag~\cite{herminghaus2014_interfacial},%
\begin{equation}
F_{S}=2\pi R_{c}v\frac{2\eta_{\scriptscriptstyle{5\mathrm{CB}}}+3\eta_{\scriptscriptstyle{aq}}}{1+\eta_{\scriptscriptstyle{aq}}/\eta_{\scriptscriptstyle{5\mathrm{CB}}}},%
\end{equation}
acting on an aqueous core moving through bulk 5CB at $v \approx 6 \,\mu$m/s, which is comparable to the velocity of the convective flow in our shells. 
We propose that the nemato-elastic repulsion provides a significant, although not insurmountable, barrier against coalescence. 

We have analysed the meandering dynamics by simultaneously tracking the circulation of the flow inside the core $\Gamma(t)$, local trajectory curvature $\kappa(t)$, and propulsion speed $V(t)$ (Fig.~\ref{fig:coreorient}\,a--c). 
In quasi-2D confinement, the core is trapped off-axis inside the convective torus, where it co-rotates with the convective flow, as shown by the core flow and colour coded $\Gamma$ values in Fig.~\ref{fig:coreorient}\,a,\,b. In this arrangement, there is less viscous resistance to the driving interfacial flows in the part of the shell containing the core, resulting in asymmetric flow with respect to the direction of motion (shown by the colour bar in Fig.~\ref{fig:coreorient}\,a), and a curved trajectory. This eventually curves the shell back towards its own trail, where chemotactic repulsion causes $V$ and $\Gamma$ to slowly decay and then abruptly reverse - the tip of the `shark-fin' motion (Fig.~\ref{fig:coreorient}\,b,\,c). This abrupt reorientation corresponds to a spike in the local curvature and is followed by a sharp acceleration (Fig.~\ref{fig:coreorient}\,c), caused by repulsion from the local gradient of filled micelles. Due to the flow reversal, in the co-moving reference frame the core has now switched sides and the shell curves in the opposite direction, once again towards its own trail. We distinguish three time scales: a short timescale ($\approx1$\,s) for autochemotactically driven abrupt reorientation, an intermediate timescale ($\approx5$\,s) for the curved motion between two shark-fin tips, and a long timescale ($>100$\,s) corresponding to the persistent motion imposed by the chemical field in the trail of the shell (\textit{cf}.~\cite{SM}, Fig.~S2).

To investigate the role of the core in breaking the flow symmetry, we have compared the 3D motion of droplets with zero, one and two cores in a density matched medium (Fig.~\ref{fig:coreorient}\,d--f). 
With no core, we reproduce previous findings~\cite{kruger2016_curling}, where the displacement of the radial `hedgehog' defect induces a torque on the droplet, resulting in helical trajectories. 
With one core, we observe similar behaviour. 
Given the freedom of a third dimension, the droplet is not arrested by its own trail and does not reverse its direction. 
Instead, the core precesses around the axis of motion. 
Shells propel in more tightly wound helices than single emulsions, which can be understood in terms of the torque applied by the respective viscous anisotropy: for shells, it is the viscosity ratio of oil and water, $\eta_2(\mathrm{5CB})/\eta(\mathrm{H_2O})\approx50$; in contrast, for single emulsions~\cite{kruger2016_curling}, it refers to the intrinsic viscous anisotropy of a nematic liquid crystal $\eta_2(5\mathrm{CB})/\eta_{iso}(5\mathrm{CB})\approx3$~\cite{chmielewski1986_viscosity}. 
With two cores, this broken symmetry argument does not hold, and thus we are able to rectify the meandering motion. 

\begin{figure}
\includegraphics[width=\columnwidth]{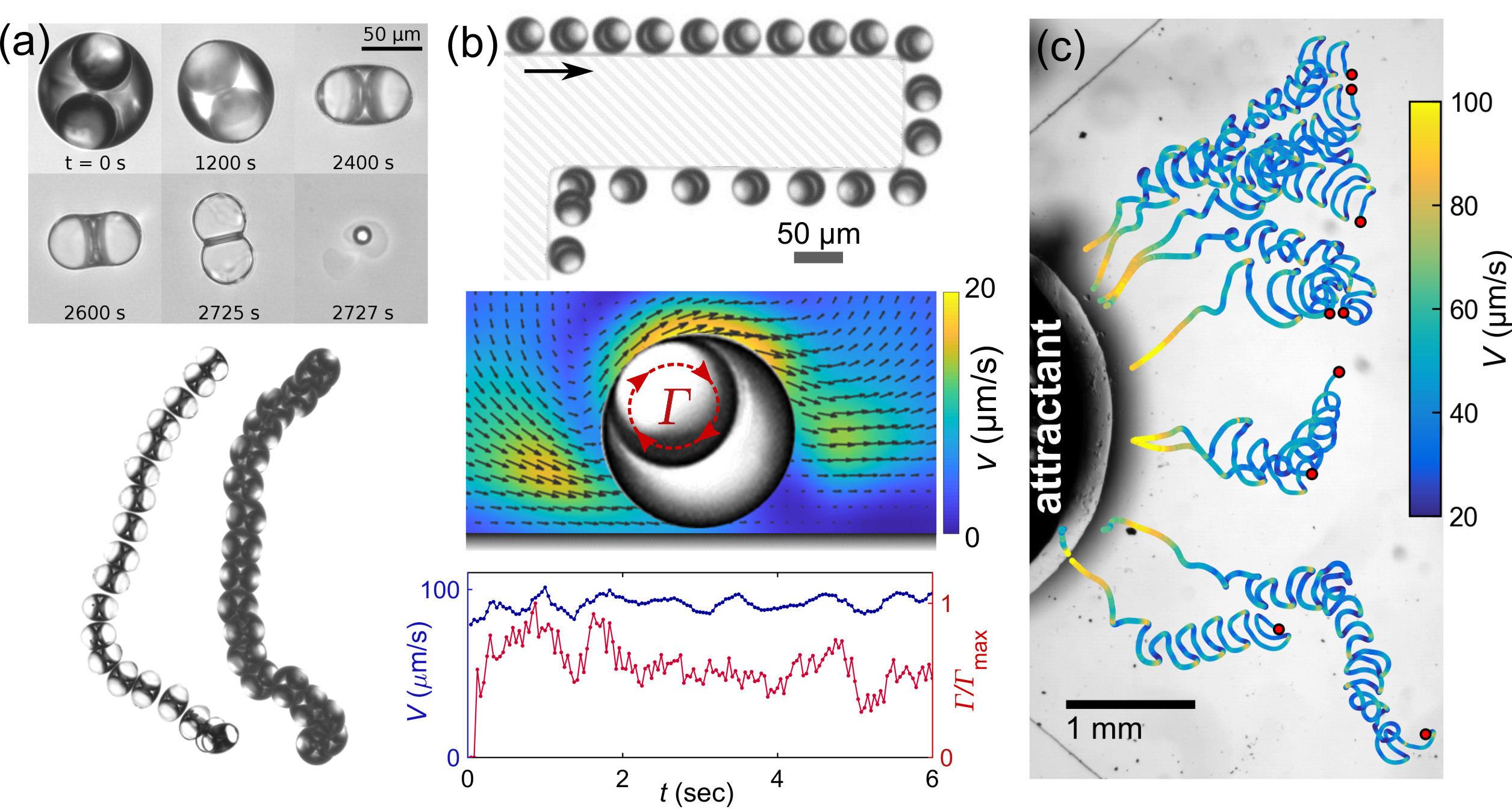}
\caption{\label{fig:guidance} Control of shell dynamics. (a) Micrographs of shells with two cores, showing long-time stability (top) and no meandering in 2D confinement (bottom, multiple exposure micrographs at two shell thicknesses).
(b) Topographical guidance by walls. Top: multiple exposure micrograph over 65\,s.
Middle: Flow field $\textbf{v}$ around the shell near a wall, mapped by PIV.
Bottom: Plot of swimming speed $V$ and core circulation $\Gamma$ when moving parallel to a wall (\textit{cf}.\ Fig.~\ref{fig:coreorient}\,c).
(c) Chemotaxis: diffusing surfactant guides shells to the left (trajectories coloured by shell speed $V$).}
\end{figure}

While a single-core nematic shell is defect free and spherically symmetric at rest, double core shells have a fixed axis set by the two cores, with a topological charge of +1 resolved by a hyperbolic hedgehog defect, or a defect loop \cite{mermin1979_topological,lubensky1998_topological,poulin1998_inverted}. 
This defect provides a barrier against core coalescence~\cite{poulin1997_novel}. 
Hence, as in the single-core case, the shell thickness shrinks to $\approx1\,\mu$m until the shell bursts (Fig.~\ref{fig:guidance}\,a). 
The most likely flow field configuration inside a moving double-core shell is with both cores trapped on opposite sides of the convection torus and no symmetry breaking mechanism or curling. 
Instead, the shell moves perpendicularly to the core alignment, with some rotational fluctuations (demonstrated in quasi-2D, Fig.~\ref{fig:guidance}\,a). 

In addition to topological methods for rectification of the propulsion dynamics, we now demonstrate two methods to guide self-propelling shells and improve their utility as cargo carrying vessels. 
First, topographical guidance: Fig.~\ref{fig:guidance}\,b shows a shell swimming along a wall~\cite{jin2018_chemotactic}, turning both convex and concave corners (further examples in Fig.~S4~\cite{SM}). 
PIV data shows that for such a shell the flow is fastest at the interface furthest from the wall, which coincides with the position of the core. 
This results in a torque on the droplet, directed towards the wall, stabilizing wall alignment. 
In contrast to the meandering case, $\Gamma$ and $V$ are constant, as the wall prevents the shell from curling towards its own trail. 
Second, chemotactic guidance: as with single emulsions~\cite{jin2017_chemotaxis}, shells follow external surfactant concentration gradients. 
In Fig.~\ref{fig:guidance}\,c, crystalline surfactant (`attractant') is allowed to dissolve into a Hele-Shaw cell. 
The resulting gradient extends $\approx$\,1\,mm into the cell, attracting the shells, doubling their speed and rectifying the meandering instability. 

In conclusion, we have developed a versatile platform for microscopic cargo delivery: self-propelling droplet shells. 
While motility induces convection that acts to destabilize these cargo vessels, we have demonstrated through experiments and simulations that nemato-elasticity can be employed as a topologically stabilizing agent, a fact we anticipate will be utilised in novel designs of microreactors and artificial cells. 
We have also provided pathways for guiding the trajectories of these droplets, through both chemical signalling and topography. 
Finally, we have analysed the interesting swimming behaviour of these self-propelling shells, and anticipate that the understanding of the rich `shark-fin' meandering dynamics will impact the design of artificial microswimmers, where swimming behaviour can be tweaked by tuning the routes for spontaneous symmetry breaking.

\cleardoublepage

\beginsupplement

\onecolumngrid 

{\begin{center} \textbf{\large Topological Stabilization and Dynamics of Self-propelling Nematic Shells: Supplemental Material}\end{center}\vspace{2mm}}

\twocolumngrid
\begin{figure*}
\centering\includegraphics[width=.8\textwidth]{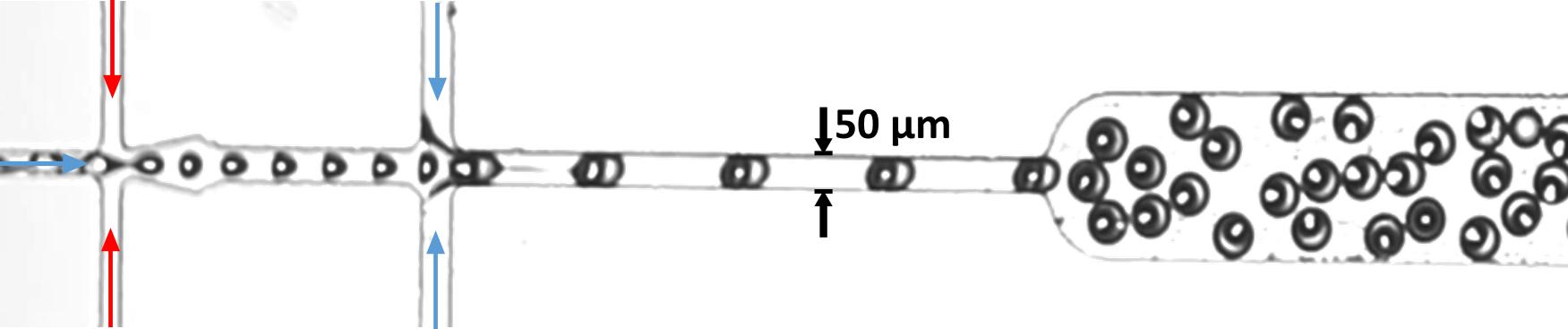}
\caption{Screenshot of a microfluidic chip producing double emulsion droplets ($R_{s}\approx 30\,\mu$m). Blue and red arrows indicate flow of aqueous and oil phases respectively. The channel height is 50\,$\mu$m.}\label{fig:doublejunction}
\end{figure*}

\subsection{Double emulsion fabrication} 
Our emulsions comprise the nematic liquid crystal 4-pentyl-4'-cyano-biphenyl (5CB) and an aqueous solution of the surfactant tetradecyltrimethylammonium bromide (TTAB). Where required, we match the densities between the oil ($\rho_{\scriptscriptstyle{5CB}} = 1.022\mathrm{{\,g\,ml^{-1}}}$ at 24$^\circ$C)
and surfactant solution phases by D$_2$O ($\rho_{\scriptscriptstyle{D_2O}}=1.107\mathrm{{\,g\,ml^{-1}}}$) admixture, with adjustments for TTAB content.
In separate experiments, we created isotropic double emulsions, using: the branched 5CB isomer CB15 ((S)-4-Cyano-4'-(2-methylbutyl)biphenyl, Synthon Chemicals); or a 10:1 volumetric ratio of 5CB and BPD (Bromopentadecane, Sigma-Aldrich). These shells were highly unstable when active, and were not examined further. 

We generate and observe double emulsion droplets in microfluidic PDMS chips on glass slides; fabricated in-house using standard soft lithography methods~\cite{qin2010_soft}.
We follow the recipe of Petit \emph{et al.}~\cite{petit2016_vesicles-on-a-chip}, for creating monodisperse water-in-oil-in-water emulsions, where the chips consist of two sequential cross-shaped flow junctions, for water-in-oil followed by oil-in-water droplet pinch-off (Fig.~\ref{fig:doublejunction}). The number of cores, as well as the core and shell diameters, are set by the flow rates~\cite{schmit2014_commensurability-driven}. Shell and core radii were produced in the range of $R_{s}=20$-$130\,\mathrm{{\mu m}}$ and $R_{c}=10$-$50\,\mathrm{{\mu m}}$ respectively. Typical flow rates were in the ranges 10-30, 60-150, and 200-500 $\mathrm{{\mu l\,hr^{-1}}}$ for the core aqueous phase, oil phase, and external aqueous phase respectively.

To ensure the oil phase did not wet the initially hydrophobic PDMS walls, the outer flow channels were hydrophilised prior to droplet production. This was achieved by drawing a sequence of liquids through the outermost channel via a vacuum pump induced pressure gradient. The sequence was as follows: a 1:1 volumetric ratio of hydrochloric acid (HCl at 37 wt.\%, Sigma-Aldrich) and hydrogen peroxide ($\mathrm{{H}_2{O}_2}$ at 30 wt.\%, Sigma-Aldrich) for 2 minutes; Milli-Q water for 30 seconds; 5 wt.\% aqueous solution of poly(diallyldimethylammonium chloride) (PDADMAC, average molecular weight ${M_w}\approx100,000$-$200,000~\mathrm{{g\,mol}^{-1}}$, Sigma-Aldrich), for 2 minutes; 2 wt.\% aqueous solution of poly(sodium 4-styrenesulphonate) (PSS, ${M_w}\approx1,000,000~\mathrm{{g\,mol}^{-1}}$, Sigma-Aldrich) for 2 minutes.

During production and in storage, both the internal and external aqueous phases contain a submicellar concentration of TTAB (0.1 wt.\%): sufficient to stabilize the interface, but insufficient to induce solubilisation ($\mathrm{CMC}=0.13$ wt.\%). 
 5CB double emulsion droplets at room temperature ($T=24^{\circ}$C) are stable against coalescence for several months.
 
\subsection{Observation and analysis}
For polarised microscopy we used a Nikon Eclipse LV100 microscope equipped with a digital camera (EOS 600d, Canon). Crossed polarisers with an added $\lambda$ retardation plate were used to visualise the nematic director field inside the liquid crystal phase. We used two inverted microscopes (Olympus IX-73 and IX-81) with either a Grasshopper (GS3-U3-41C6M-C)  greyscale camera ($2048\times2048$ pixels) or a Canon (EOS 600d) digital camera ($1920\times1080$ pixels) for regular visible light microscopy recording images at 4-24 fps and 4-$40\times$ magnification for tracking the swimmers. The fluorescent imaging was done on an IX-73 inverted microscope at 24 fps and $20\times$ magnification for PIV measurements around the shells and at 48 fps and $40\times$ magnification for PIV inside the core. We extracted shell positions from video microscopy data using standard Python libraries for numpy, PIL and opencv (Scripts available on request). Essential steps are background correction, binarisation, blob detection by contour analysis and minimum enclosing circle fits. We calculated trajectories and speeds using a simple next neighbour algorithm~\cite{crocker1996_methods}. We identified bursting times by frame-by-frame inspection. We estimated core position and orientation angles from high resolution video data by semi-automatically determining circular shell and core outlines and comparing the line of centers to the trajectory normal. The mean squared displacement of the shell trajectory was calculated as
\begin{equation}\tag{S1}
\langle(\Delta r)^2\rangle_t= \langle[\textbf{r}(t_0 + t)-\textbf{r}(t_0)]^2 \rangle_{t_0},%
\end{equation}
where \textbf{r} is the position vector of the shell. The angular or velocity autocorrelation function was calculated as
\begin{equation}\tag{S2}
C(t) = \bigg\langle \frac{\textbf{V}(t_0 + t) . \textbf{V}(t_0)}{|\textbf{V}(t_0 + t)| |\textbf{V}(t_0)|}\bigg\rangle_{t_0},
\end{equation}
and the (signed) local curvature of the trajectory was defined as 
\begin{equation}\tag{S3}
\kappa = \frac{x'y''-y'x''}{({x'}^2 + {y'}^2)^{3/2}},
\end{equation}
where the prime denotes derivation with respect to time.

\begin{figure*}
\includegraphics[width=\textwidth]{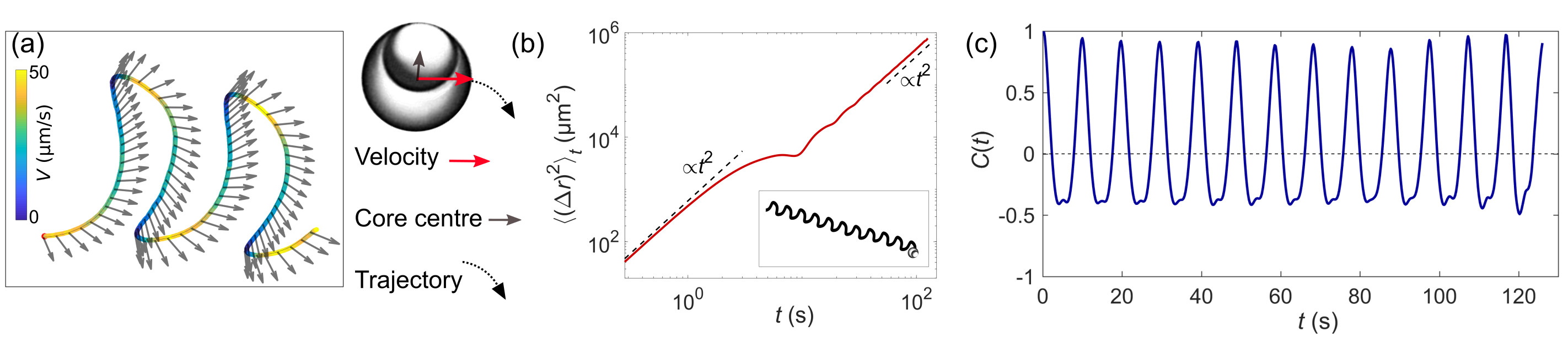}
\caption{Characterisation of the swimming behaviour. (a) The magnitude of the swimming velocity. The vectors show the location of the core at each time. The sudden acceleration right after the abrupt reorientation corresponds to the autochemotactic repulsion from the trail of filled micelles. (b) Mean squared displacement of an individual swimmer with the trajectory as the inset. The dip corresponds to the meandering period. (c) Velocity auto-correlation function of the same swimmer.\label{fig:Swimming_charac}}
\end{figure*}

For illustration, we have included speed, orientation, mean squared displacement $\langle(\Delta r)^2\rangle_t$ and velocity autocorrelation $C(t)$ data in Fig.~\ref{fig:Swimming_charac} for one trajectory (see also Fig.~3 in the main text). The periodic meandering reflects in a dip in the $\langle(\Delta r)^2\rangle_t$, which is $\propto t^2$ for long times, approaching ballistic motion. The absence of rotational noise reflects in a periodic, non-decaying character of  $C(t)$.

For the PIV analyses in Fig.~3\,a and Fig.~4\,b. the shell was pinned in a shallow cell and the aqueous phase was seeded with 0.5\,$\mu$m fluorescent tracer colloids (Thermo Fisher Scientific). For PIV inside the core, the core was seeded with the same tracers, the core position was tracked manually and the time-resolved PIV images were collected. All PIV analyses and flow calculations were performed via the MATLAB based PIVlab interface~\cite{thielicke2014_pivlab}. Additional custom MATLAB scripts were written to calculate the circulation of the flow inside the internal aqueous droplet~\cite{wioland2016_ferromagnetic}:
\begin{equation}\tag{S4}
\Gamma_i(t)= \hat{\textbf{ z }} . \bigg [\sum\nolimits_{(x,y)_i} \textit{\textbf{r}}_i(x,y) \times \textbf{u}(x,y,t) \bigg ],%
\end{equation}
where $\textit{\textbf{r}}_i$ is the vector from the core center to $(x,y)$ inside the projected area of the core on the x-y plane and $\textbf{u}$ is the velocity vector field inside the core.

\subsection{Numerical simulation}
The estimated elastic energies ($E$) plotted in Fig.~2\,b, c were calculated by applying a common numeric minimisation technique~\cite{mori1999_multidimensional,ravnik2009_landau-de} based on the $\mathbf{Q}$ tensor representation~\cite{degennes1971_short} of the nematic director field, where we have assumed that the 3 nematic elasticity constants are equal, and the scalar order parameter is equal to 1 in a defect-free shell. More details of these calculations have been included in the main manuscript.

\begin{figure*}
\includegraphics[width=\textwidth]{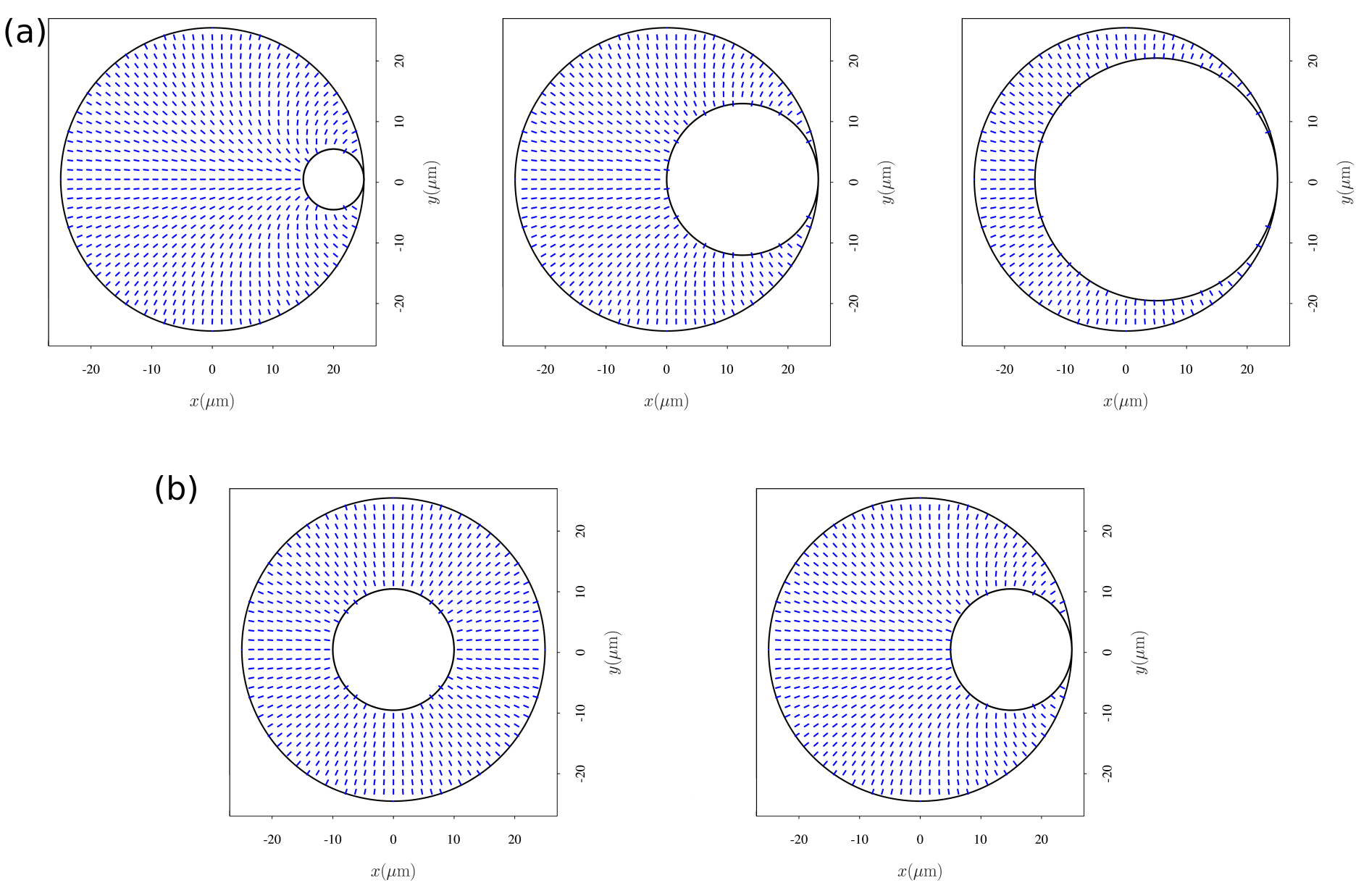}
\caption{Simulation of the director field. Row (a) shows the effect of size ratio for the maximum displacement of the core from the center, row (b) shows the uniform (left) and distorted director fields (right), indicating the effect of core displacement on the deviation of the director field.}\label{fig:DirectorField}
\end{figure*}

In order to determine the structure possessing the minimal value of $E$, the shell was mapped onto a cubic grid with $256^3$ nodes. The tensor components at each node were assumed to relax to their equilibrium values according to a simplified equation of motion, governed only by the rotational viscosity $\gamma_1$. In a discretised form with time steps $\Delta t$ we have:
\begin{equation}\tag{S5}
\gamma_1 \frac{\Delta Q_{jk}}{\Delta t} = - \frac{\delta f_e}{\delta Q_{jk}}.
\end{equation}
The right-hand side of the above equation represents the functional derivative of $f_e$ with respect to tensor component $Q_{jk}$. Details of the numerical procedures and explicit expressions for the various functional derivatives are given in~\cite{mori1999_multidimensional}. For the boundary conditions we assumed strict homeotropic anchoring of the director at all interfaces, \emph{i.e.}, for the nodes located at the outer or inner surface of the shell, the orientation of the nematic director at the interface is set perpendicular to the interface and held constant (so-called strong anchoring conditions). We varied the shell size and the position of the inner aqueous droplet over a wide range and determined the corresponding equilibrium structures and values of $E$. Fig.~\ref{fig:DirectorField} illustrates some examples. We note that we considered here solely the interplay between the shell size, the position of the internal core, and the elastic energy of the nematic director field. We neglected the internal convectional flow which is present in our self-propelling shells, which certainly leads to additional effects on the structure of the director field.

\begin{figure*}
\includegraphics[width=.7\textwidth]{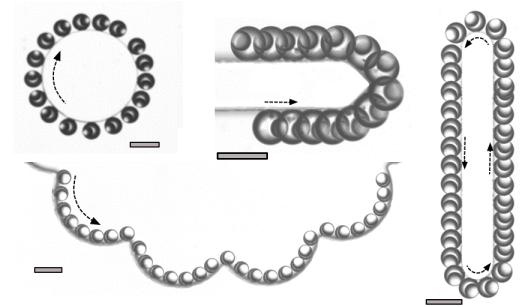}
\caption{Further examples of topographical guidance of the shells, including around edges and corners, \emph{cf.} Fig.~4, main text. The scale bar is 100\,$\mu$m.\label{fig:Topography}}
\end{figure*}

\cleardoublepage
 \begin{figure*}[t]
Supplementary movies are available for viewing under http://asm.ds.mpg.de/index.php/media/.\\\vspace{1cm}%
\begin{minipage}{.49\textwidth}
\includegraphics[width=\columnwidth]{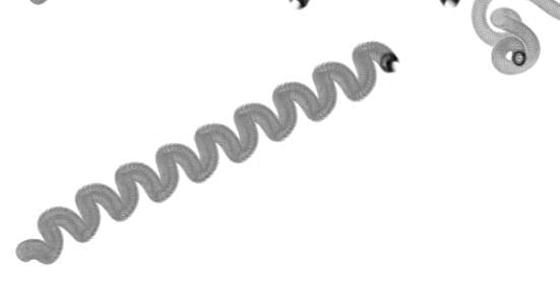}
\caption{Movie S5 showing one shell  ($R_{s}\approx 30\,\mu$m, 131\,s, speed x\,6 at 24\,fps) `shark-fin' meandering, with high persistence and periodicity, as well as a `curling' single emulsion droplet, in a uniform bulk environment.\label{fig:movie2}}\vspace{5mm}
\includegraphics[width=\columnwidth]{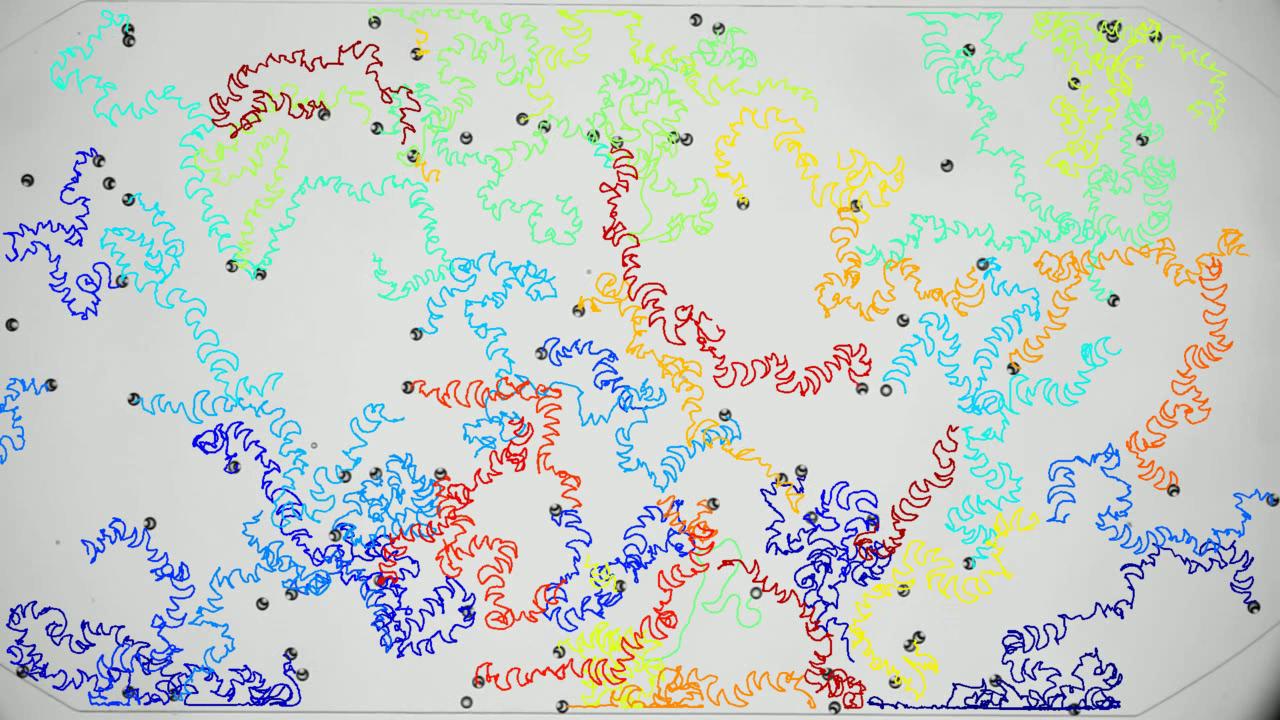}
\caption{Movie S6 showing selected trajectories for an ensemble of `shark-fin' meandering shells (500\,s, speed x\,25 at 10\,fps, field of view 5.4\,x\,3\,mm). Reorientation on long timescales is caused by the filled micelles in the wake of previously passing shells.\label{fig:movie3}}
\end{minipage}\hspace{.\stretch{1}}%
\begin{minipage}{.49\textwidth}
\includegraphics[width=\columnwidth]{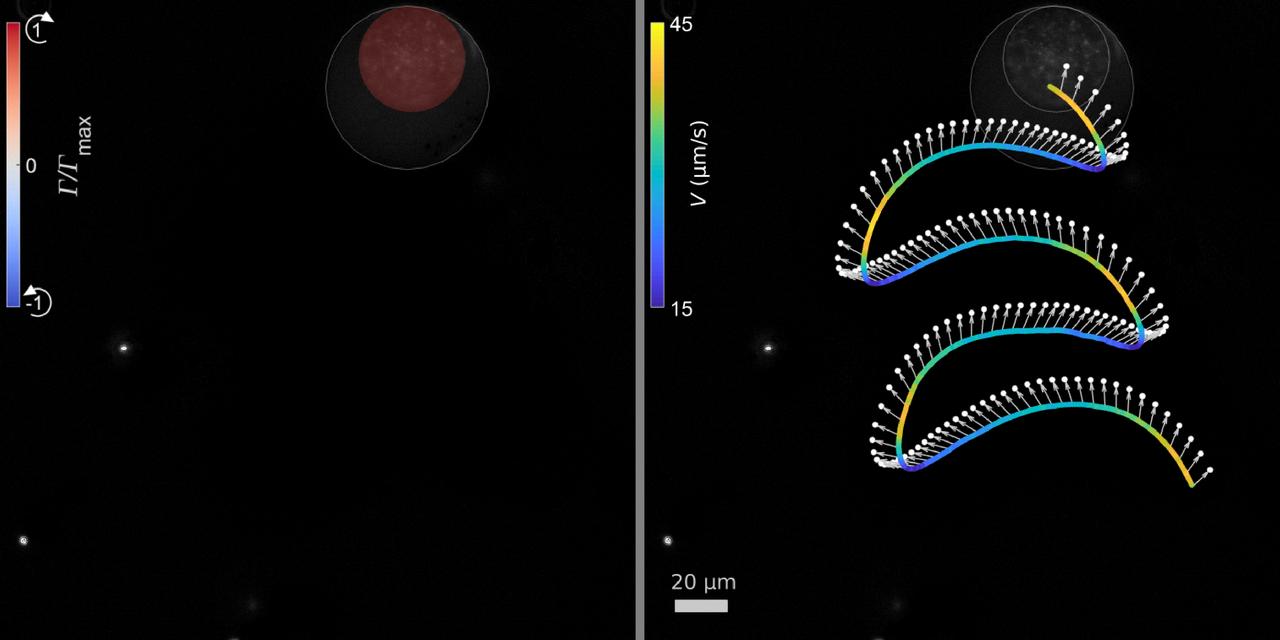}
\caption{Movie S7 (255\,s, speed x\,1 at 24\,fps) tracking circulation, speed and core orientation for a meandering shell, data corresponding to Fig.~3\, b,\,c in the main text.\label{fig:movie5}}\vspace{5mm}
\centering\includegraphics[width=.8\columnwidth]{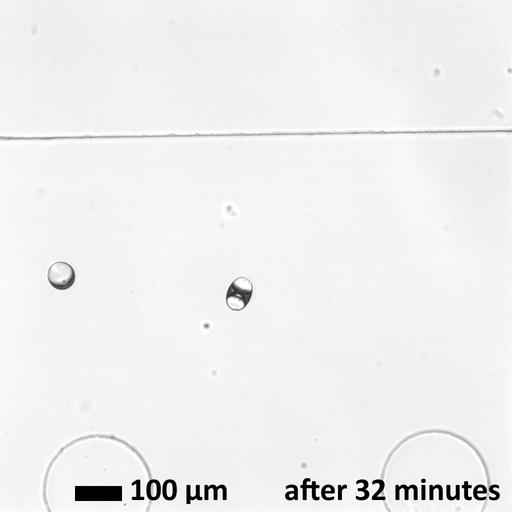}
\caption{Movie S8 (speed x\,12 at 24\,fps) showing the dynamics and life stages of a double core shell (see Fig.~4 in the main text).\label{fig:movie4}}
 \end{minipage}
\end{figure*}

\end{document}